\documentclass[prl,showpacs,preprintnumbers]{revtex4}
\usepackage{epsfig}
\usepackage{amssymb,amsmath,amsthm}

\begin{document}

\newcommand{\hdblarrow}{H\makebox[0.9ex][l]{$\downdownarrows$}-}
\title{Modeling Kelvin Wave Cascades in Superfluid Helium}

\author{Guido ~Boffetta$^{1}$, Antonio ~Celani$^{2,1}$, Davide ~Dezzani$^{1}$,
Jason Laurie$^3$ and Sergey Nazarenko$^3$}

\address{$^1$ Dipartimento di Fisica Generale and INFN,
Universit\`a degli Studi di Torino, v. Pietro Giuria 1, 10125, Torino, Italy \\
and CNR-ISAC, Sezione di Torino, c. Fiume 4, 10133 Torino, Italy \\
$^2$ CNRS, Institut Pasteur, Rue du docteur Roux 25, 75015 Paris, France\\
$^3$ Mathematics Institute, University of Warwick, Coventry CV4 7AL, UK}

\date{\today}


\begin{abstract}
We study two different types of simplified models for Kelvin wave turbulence on quantized vortex lines in superfluids near zero temperature.
Our first model is obtained from a truncated expansion of the  Local Induction Approximation (Truncated-LIA)
and it is shown to possess the same scalings and the essential behaviour as the
full Biot-Savart model, being much simpler than the latter and, therefore, more
amenable to theoretical and numerical investigations.
The Truncated-LIA model supports six-wave interactions and dual cascades, which are clearly demonstrated via the direct numerical simulation of this model in the present paper. In particular, our simulations confirm presence of the weak turbulence
regime and the theoretically predicted spectra for the direct energy cascade and the inverse wave action cascade.
The second type of model we study, the Differential Approximation Model (DAM), takes a further drastic simplification by assuming locality
of interactions in $k$-space via a differential closure that preserves the
main scalings
of the Kelvin wave dynamics.  DAMs are even more amenable to study and they form a useful tool by providing simple
analytical solutions in the cases when extra physical effects are present, e.g. forcing by reconnections, friction dissipation and
phonon radiation. We study  these models numerically and test their theoretical predictions, in particular the formation of the
stationary spectra, and the closeness of the numerics for the higher-order DAM
to the analytical predictions for the lower-order DAM .

\end{abstract}


\maketitle

\section{Introduction}
It is well known that a classical vortex filament
can support linear waves. These
were predicted by Kelvin more than one century ago and experimentally
observed about 50 years ago in superfluid $^{4}He$.
At very low temperature, where the friction induced by normal fluid
component can be neglected, Kelvin waves can be dissipated only at
very high frequencies by phonon emission \cite{V01}.
Therefore at lower frequency, energy is transferred among different
wavenumbers by nonlinear coupling.
This is the mechanism at the basis of the Kelvin wave
cascade which sustains superfluid turbulence \cite{S95,V00}.

In recent years, the single vortex Kelvin wave cascade has attracted
much theoretical \cite{KS04,N06}, numerical \cite{KVSB01,VTM03,KS05} and
experimental \cite{WGHLV07} attention.
Even within the classical one-dimensional vortex model, different
degrees of simplification are possible. For small amplitudes, the
vortex configuration can be described by a two component vector field, made of
the coordinates of the vortex line in the plane transverse to the
direction of the unperturbed filament. These depend on the single coordinate
that runs
along the filament.
As was shown in \cite{S95}, this system of equations admits a Hamiltonian
formulation, dubbed the two-dimensional Biot-Savart formulation (2D-BS), see
(2) below.
Another, more drastic, simplification is obtained by considering local
interactions only. This leads to the local induction approximation
(LIA) which was originally derived starting from the full 3D-BS
\cite{AH65}.
The main limitation of LIA is that it generates an integrable system
with infinite conserved quantities, as it is equivalent to the nonlinear
Schr\"odinger equation \cite{Hasimoto72}, and therefore, the resonant wave interactions are
absent  (at all orders) and one cannot
 reproduce the phenomenology of the
full system.  For this reason LIA, despite its simplicity, is of
little help for the study of {\it weak} Kelvin wave turbulence.
On the other hand, LIA contains solutions leading to self-crossings (numerical  \cite{Schwarz}
and analytical  \cite{S95}) and, therefore, it
can qualitatively describe vortex line reconnections in {\em strong} 3D-BS turbulence (``vortex tangle").


In this paper we consider simple models for a vortex
filament that is able to sustain a turbulent energy cascade. The first model
is obtained in the limit of small amplitudes by a Taylor expansion
of the 2D-LIA. The truncation breaks the integrability of the Hamiltonian
and therefore generates a dynamical system with two inviscid
invariants (energy and wave action). For this class of systems,
whose prototype is the two-dimensional Navier-Stokes
turbulence \cite{KM80}, we expect a dual cascade phenomenology
in which one quantity flows to small scales generating a
{\it direct cascade} while the other goes to larger scales
producing an {\it inverse cascade}. The possibility of a dual cascade
scenario for Kelvin waves turbulence has been recently suggested
\cite{lebedev,N06} but never observed, and in this paper we present the first
numerical evidence for the inverse cascade.

The second class of simplified models, Differential Approximation Models (DAMs),
use a closure in which the multi-dimensional $k$-space integral in the wave interaction term
(collision integral in the wave kinetic equation) is replaced by a nonlinear
differential term  that preserve the main properties and  scalings
of the Kelvin wave dynamics such as the energy and wave action conservations,
scaling of the characteristic evolution time with respect to the wave intensity and the
wavenumber $k$. DAMs have proved to be a very useful tool in the analysis of fluid
dynamical and wave turbulence in the past \cite{L68,H85,I85,ZP99,c04,L04,N06,L06},
and here we study them in the context of the Kelvin wave turbulence.
DAMs are particularly useful when one would like to understand the temporal evolution
of the spectrum, when the physical forcing and dissipation need to be included, or
when the Kelvin wave system is subject to more involved boundary conditions
leading to simultaneous presence of two cascades in the same range of scales,
or a thermalization (bottleneck) spectrum accumulation near a flux-reflecting
boundary . In the second part of this paper, we will present numerical studies
of DAMs in presence of some of these physical factors
and we will test some previously obtained analytical predictions.

\section{BSE}
At a macroscopic level, the superfluid vortex filament is a classical
object whose dynamics is often described by the Biot-Savart equation (BSE)
\begin{equation}
\dot{\bf r} = {\kappa \over 4 \pi} \int {d {\bf s} \times ({\bf r}-{\bf s})
\over |{\bf r}-{\bf s}|^3},
\label{eq:bse}
\end{equation}
which describes the self-interaction of vortex elements.
The quantum nature of the phenomenon
is encoded in the discreteness of circulation $\kappa=h/m$ \cite{Donnelly91}.

The BSE dynamics of the vortex filament admits a Hamiltonian formulation
under a simple geometrical constraint: the position ${\bf r}$ of
the vortex is represented in a two-dimensional parametric form
as ${\bf r}=(x(z),y(z),z)$, where $z$ is a given axis.
From a geometrical point of view, this corresponds to small
perturbations with respect to the straight line configuration, i.e.
the vortex cannot form folds in order to preserve the single-valuedness
of the $x$ and $y$ functions.
In terms of the complex canonical coordinate $w(z,t)=x(z,t)+i y(z,t)$,
the BSE can be written in a Hamiltonian form $i \dot{w} = \delta H[w]/\delta w^*$
with \cite{S95}
\begin{equation}
H[w] = {\kappa \over 4 \pi} \int dz_1 dz_2 {1 + Re(w'^{*}(z_1) w'(z_2))
\over \sqrt{(z_1 - z_2)^2 + |w(z_1)-w(z_2)|^2}},
\label{eq:h2d}
\end{equation}
where we have used the notation $w'(z)=d w/ d z$.
The geometrical constraint of a small amplitude
perturbation can be expressed in terms of a parameter
$\epsilon(z_1,z_2)=|w(z_1)-w(z_2)|/|z_1 - z_2| \ll 1$.

An enormous simplification, both for theoretical and numerical
purposes, is obtained by means of the so called local induction approximation
(LIA) \cite{AH65}. This approximation is justified by the observation
that (\ref{eq:bse}) is divergent as ${\bf s} \to {\bf r}$ and
is obtained by introducing a cutoff at $a<|{\bf r}-{\bf s}|$
in the integral in (\ref{eq:bse}) which represents
the vortex filament radius.

When applied to Hamiltonian (\ref{eq:h2d}), the LIA procedure gives \cite{S95}
\begin{equation}
H[w] = 2 {\kappa \over 4 \pi} \ln\left({\ell \over a}\right)
\int dz \sqrt{1 + |w'(z)|^2}
= 2 \beta L[w],
\label{eq:hlia}
\end{equation}
where $\ell$ is a length of the order of the
curvature radius (or inter-vortex distance when the considered vortex filament is a
part a vortex tangle),
$\beta=(\kappa/4 \pi)\ln(\ell/a)$. Here, it was taken into account that because
$a$ is much smaller than any other characteristic size in the system,
$\beta$ will be about the same whatever characteristic scale $\ell$ we take in its definition.
We remark that in the LIA approximation the Hamiltonian is proportional
to the vortex length $L[w]=\int dz \sqrt{1 + |w'(z)|^2}$ which is
therefore a conserved quantity.
The equation of motion from (\ref{eq:hlia}) is (we set $\beta=1/2$
without loss of generality, i.e. we rescale
time as $2 \beta t \to t$)
\begin{equation}
\dot{w} = {i \over 2} \left( {w' \over \sqrt{1+|w'|^2}} \right)'.
\label{eq:lia}
\end{equation}
As a consequence of the invariance under phase transformations,
equation (\ref{eq:lia}) also conserves the total wave action
(also called the kelvon number \cite{KS04})
\begin{equation}
N[w] = \int dz |w|^2.
\label{eq:n}
\end{equation}

In addition to these two conserved quantities, the 2D-LIA model
possesses an infinite set of invariants and is
integrable, as it is the LIA of BSE (which can be transformed into
the nonlinear Schr\"odinger equation by the Hasimoto transformation, see Appendix B).
Due to the integrability, in weak Kelvin wave turbulence~\footnote{Weakness of waves
implies that there are no vortex line reconnections. On the other hand, for strong waves
the reconnections can occur and they could qualitatively be described by the self-crossing
solutions of LIA \cite{S95}. At self-crossing events, the LIA model fails, but it can be ``reset"
via an {\em ad hoc} reconnection procedure \cite{Schwarz}.}
the energy and the wave action cannot cascade within the LIA
model, but this can be fixed by a simple truncation as we show in the next section.


\section{Truncated LIA}
Integrability is broken if one considers a truncated expansion of
the Hamiltonian (\ref{eq:hlia}) in power of wave amplitude $w'(z)$.
Taking into account the lower order terms only, one obtains:
\begin{eqnarray}
H_{exp}[w] &=& H_0 + H_1 + H_2 =  \nonumber \\
&=& \int dz \left( 1 + {1 \over 2} |w'|^2 - {1 \over 8} |w'|^4 \right).
\label{eq:expansion}
\end{eqnarray}
Neglecting the constant term, the Hamiltonian can be written in
Fourier space as
\begin{eqnarray}
H_{exp}&=&\int \omega_k |w_k|^2 dk + \int dk_{1234} W_{1234} \delta^{12}_{34} w_1^* w_2^* w_3 w_4,
\label{eq:expfourier}
\end{eqnarray}
with $\omega=k^2/2$, $W_{1234}=-\frac{1}{8} k_1 k_2 k_3 k_4$ and
we used the standard notation
$\delta^{12}_{34}=\delta(k_1+k_2-k_3-k_4)$ and
$dk_{1234}=dk_1 dk_2 dk_3 dk_4$.

In Wave Turbulence, the near-identity transformation allows one
to eliminate ``unnecessary" lower orders of nonlinearity in the system
if the corresponding order of the wave interaction process is nil \cite{ZLF92}.
For example, if there are no three-wave resonances, then one can eliminate
the cubic Hamiltonian.
(The quadratic Hamiltonian corresponding to the linear dynamics, of course, stays).
 This process can be repeated recursively, in a way similar to the KAM theory, until
 the lowest order of the non-trivial resonances is reached. If no such resonances appear
 in any order, one has an integrable system.

 In our case, there are no four-wave resonances (there are no non-trivial solution
 for the resonance conditions for $\omega \sim k^x$ if $x>1$ in one dimension). However, there are nontrivial solutions of the six-wave resonant
 conditions. Thus, one can use the near-identity transformation to convert system (\ref{eq:expfourier}) into
 the one with the lowest order nonlinear interaction to be of degree six, (there are no five-wave resonances, the interaction coefficients in the quintic Hamiltonian are identically equal to zero after applying the canonical transformation.)

 A trick for finding a shortcut derivation of such a transformation is described in
 \cite{ZLF92}. It relies on the fact that the time evolution operator is a canonical
 transformation.
   Taking the Taylor expansion of $w(k,t)$ around
$w(k,0)=c(k,0)$ we get a desired transformation, that is by its
derivation, canonical.  The coefficients of each term can be
calculated from an auxiliary Hamiltonian $H_{aux}$,

\begin{eqnarray}
H_{aux}&=&\int \tilde{V}_{123}\delta^{12}_3\left(c_1c_2c_3^* + \mathrm{c.c.}\right) dk_{123}\nonumber\\
&&+\int \tilde{U}_{123}\delta^{123}\left(c_1c_2c_3 + \mathrm{c.c.}\right) dk_{123}\nonumber\\
&&+\int \tilde{W}_{1234}\delta^{12}_{34}c^*_1c^*_2c_3c_4 dk_{1234}\nonumber\\
&&+\int \tilde{X}_{1234}\delta^{1}_{234}\left(c_1c_2^*c_3^*c_4^* + \mathrm{c.c.}\right) dk_{1234}\nonumber\\
&&+\int \tilde{Y}_{1234}\delta^{1234}\left(c_1c_2c_3c_4  + \mathrm{c.c.}\right) dk_{1234}\nonumber\\
&&+\int \tilde{Z}_{12345}\delta^{12345}\left(c_1c_2c_3c_4c_5  + \mathrm{c.c.}\right) dk_{12345}\nonumber\\
&&+\int \tilde{A}_{12345}\delta^{1}_{2345}\left(c_1c_2^*c_3^*c_4^*c_5^*  + \mathrm{c.c.}\right) dk_{12345}\nonumber\\
&&+\int \tilde{B}_{12345}\delta^{12}_{345}\left(c_1c_2c_3^*c_4^*c_5^*  + \mathrm{c.c.}\right) dk_{12345}\nonumber\\
&&+\int \tilde{C}_{123456}\delta^{123}_{456}c^*_1c^*_2c^*_3c_4c_5c_6 dk_{123456}.\nonumber\\
\end{eqnarray}

The auxiliary Hamiltonian $H_{aux}$ represents a generic Hamiltonian for the canonical variable $c_k$, thus, defining $c_k$ in the canonical transformation will set the interaction coefficients of $H_{aux}$.
Here all interaction coefficients, (terms denoted with tildes) present in the auxiliary Hamiltonian are arbitrary. A similar procedure was
done in Appendix A$3$ of \cite{ZLF92} to eliminate the cubic Hamiltonian in cases when the three-wave
interaction is nil, and here we apply a similar approach to eliminate the quadric Hamiltonian.
 The transformation is represented as
\begin{equation}\label{eq:trans}
w_k=c(k,0)+t\left(\frac{\partial c(k,t)}{\partial
t}\right)_{t=0}+\frac{t^2}{2}\left(\frac{\partial^2 c(k,t)}{\partial
t^2}\right)_{t=0}+\cdots.
\end{equation}
The transformation is canonical for all $t$, so for simplicity we set $t=1$. The coefficients of (\ref{eq:trans}) can be calculated from the following formulae,

\begin{eqnarray}\label{eq:HamSys}
\left(\frac{\partial c(k,t)}{\partial
t}\right)_{t=0}&=&-i\frac{\delta H_{aux}}{\delta c^*},\nonumber \\
 \left(\frac{\partial^2c(k,t)}{\partial t^2}\right)_{t=0}&=&-i
\frac{\partial}{\partial t}\frac{\delta H_{aux}}{\delta c^*}.
\end{eqnarray}

Due to the original Hamiltonian (\ref{eq:expansion}) $H_{exp}$ having $U(1)$ gauge symmetry, we have no cubic order Hamiltonian terms, this greatly simplifies the canonical transformation (\ref{eq:trans}), because the absence of any non-zero three-wave interaction coefficients in $H_{exp}$ automatically fixes the arbitrary cubic (and quintic) interaction coefficients within the auxiliary Hamiltonian $H_{aux}$ to zero. Thus, transformation (\ref{eq:trans}) reduces to

\begin{eqnarray}\label{eq:trans2}
w_k&=&c_k-\frac{i}{2}\int d k_{234}\tilde{W}_{k234}c^*_2c_3c_4\delta^{k2}_{34}-3i\int d
k_{23456}\tilde{C}_{k23456}\delta^{k23}_{456}c^*_2c^*_3c_4c_5c_6\nonumber\\
&&+\frac{1}{8}\int d k_{234567}\Big(\tilde{W}_{k743}\tilde{W}^*_{7623}\delta^{k7}_{45}\delta^{76}_{23}-2\tilde{W}_{k247}\tilde{W}_{7356}\delta^{k2}_{47}\delta^{37}_{56}\Big)c^*_2c^*_3c_4c_5c_6 .
\end{eqnarray}

To eliminate the nonresonant four-wave interactions in Hamiltonian (\ref{eq:expfourier}), we substitute transformation (\ref{eq:trans2}) into Hamiltonian (\ref{eq:expfourier}).  This yields a new representation of Hamiltonian (\ref{eq:expfourier}) in variable $c_k$,  where the nonresonant terms (more specifically the four-wave interaction terms) will involve both $W_{1234}$ and $\tilde{W}_{1234}$.  Arbitrariness of $\tilde{W}_{1234}$ enables us to select this to eliminate the total four-wave interaction term, $H_2$.  In our case this selection is,
\begin{equation}
\tilde{W}^*_{1234}=\frac{4iW_{1234}}{\omega_1+\omega_2-\omega_3-\omega_4}.
\end{equation}
This choice is valid as the denominator will not vanish due to the nonresonance of four-wave interactions.  Hamiltonian (\ref{eq:expfourier}) expressed in variable $c_k$, $H_{c}$ becomes

\begin{eqnarray}\label{eq:fullham}
H_{c}&&=\int \omega_k c_kc_k^* d
k\nonumber\\
&&+\int
\Big(C_{123456}-i\left(\omega_1+\omega_2+\omega_3-\omega_4-\omega_5-\omega_6\right)\nonumber\\
&&\times\tilde{C}_{123456}\Big)\delta^{123}_{456}c^*_1c^*_2c^*_3c_4c_5c_6 d
k_{123456}.
\end{eqnarray}

$\tilde{C}_{123456}$ is the arbitrary six-wave interaction coefficient arising from the auxiliary Hamiltonian.  This term does not contribute to the six-wave resonant dynamics as the factor in front will vanish upon the resonant manifold, that appears in the kinetic equation.  $C_{123456}$ is the six-wave interaction coefficient resulting from the canonical transformation which is defined later in equation (\ref{eq:C123456}).

To deal with the arbitrary interaction coefficient $\tilde{C}_{123456}$, one can decompose $C_{123456}$ into its value taken on the six-wave resonant manifold plus the residue (i.e.  $C_{123456}=-\frac{1}{16}k_1k_2k_3k_4k_5k_6 + C^*_{123456}$).  We can then choose $\tilde{C}_{123456}=-iC^*_{123456}/(\omega_1+\omega_2+\omega_3-\omega_4-\omega_5-\omega_6)$, hence, allowing the arbitrary six-wave interaction coefficient in $H_{aux}$ to directly cancel with the residual value of $C_{123456}$.  This enables us to write Hamiltonian $H_{c}$ as

\begin{equation}
 H_c=\int \omega_k |c_k|^2 dk + \nonumber  \int dk_{123456} C_{123456}\delta^{123}_{456} c_1^* c_2^* c_3^* c_4 c_5 c_6,
\label{eq:hc}
\end{equation}

The explicit form of the interaction coefficient $C_{123456}$ can be expressed by

\begin{eqnarray}
C_{123456}=-\frac{1}{18}\displaystyle\sum^{3}_{i,j,k=1, i\neq j\neq k\neq i}\displaystyle\sum^{6}_{p,q,r=4, p\neq q \neq r\neq p}&\frac{W_{p+q-iipq}W_{j+k-rrjk}}{\left(\omega_{j+k-r}+\omega_r-\omega_j-\omega_k\right)}\nonumber\\
&+\frac{W_{i+j-ppij}W_{q+r-kkqr}}{\left(\omega_{q+r-k}+\omega_k-\omega_q-\omega_r\right)}.\label{eq:C123456}
\end{eqnarray}

Zakharov and Schulman discovered a parametrisation \cite{ZS82} for the six-wave resonant condition with $\omega_k \sim k^2$,
\begin{eqnarray}
k_1&=&P+R\left[u+\frac{1}{u}-\frac{1}{v}+3v\right]\nonumber,\\
k_2&=&P+R\left[u+\frac{1}{u}+\frac{1}{v}-3v\right]\nonumber,\\
k_3&=&P -\frac{2R}{u}-2Ru,\\
k_4&=&P +\frac{2R}{u}-2Ru\nonumber,\\
k_5&=&P+R\left[u-\frac{1}{u}+\frac{1}{v}+3v\right]\nonumber,\\
k_6&=&P+R\left[u-\frac{1}{u}-\frac{1}{v}-3v\right]\nonumber.
\end{eqnarray}
This parametrisation allows us to explicitly calculate $C_{123456}$ upon the resonant manifold.  This is important because the wave kinetics take place on this manifold, that corresponds to the delta functions of wavenumbers $k$ and frequencies $\omega_k$ within the kinetic equation.
When this parametrisation is used with equation (\ref{eq:C123456}) and $W_{1234}=-\frac{1}{8}k_1k_2k_3k_4$ we find that the resonant six-wave interaction coefficient simplifies to $C_{123456}=-\frac{1}{16}k_1k_2k_3k_4k_5k_6$.  Note, that this is indeed the identical to the next term, $H_3$ in the LIA expansion with opposite sign.

 This six-order interaction coefficient
 is obtained from coupling of two fourth-order vertices of $H_2$.
It is not surprising that the resulting expression coincides, with the opposite sign,
with the interaction coefficient of $H_3$ in
(\ref{eq:hlia}). Indeed, (\ref{eq:hlia}) is an integrable model which implies that
if we retained the next order too, i.e. $H_3$, then the resulting six-wave process would
be nil, and the leading order would be an eight-wave process in this case.
In fact, in the coordinate space the Hamiltonian
is simply
\begin{eqnarray}
H_{exp}[w] &=& H_0  + H_3 =  \nonumber \\
&=& \int dz \left( {1 \over 2} |w'|^2 - {1 \over 16} |w'|^6 \right).
\label{eq:expansion-tr}
\end{eqnarray}
Thus, the existence of the six-wave process is a
consequence of the truncation (\ref{eq:expansion}) of the Hamiltonian.

Hamiltonian (\ref{eq:expfourier}) (or equivalently (\ref{eq:hc}) or (\ref{eq:expansion-tr})) constitutes the
truncated-LIA  model for Kelvin wave turbulence.
It possesses the same scaling properties as the BSE system: it conserves the
energy and the wave action, and gives rise to a dual-cascade six-wave system with an interaction
coefficient with the same order of homogeneity as the one of the BSE.
A slight further modification should be made in the time re-scaling factor as
$\beta=(\kappa/4 \pi)$, - i.e. by dropping the large log factor from the original
definition.

Physical insight into Kelvin wave turbulence is obtained
from the wave turbulence (WT) approach which yields
a kinetic equation which describes the dynamics of
the wave action density $n_k=\langle |c_k|^2 \rangle$.

The dynamical equation for the variable $c_k$ can be derived from Hamiltonian (\ref{eq:hc}) by the relation $i\partial c_k / \partial t = \delta H_c / \delta c^*_k$, and is
\begin{equation}\label{eq:cdot}
	i\frac{\partial c_k}{\partial t} - \omega_k c_k = \int dk_{23456}C_{k23456}c^*_2c^*_3c_4c_5c_6\delta^{k23}_{456}.
\end{equation}

Multiplying equation (\ref{eq:cdot}) by $c_k^*$, subtracting the complex conjugate and averaging we arrive at
\begin{equation}\label{eq:n_k}
\frac{\partial \langle c_kc_k^*\rangle}{\partial t}= 6Im\left(\int
dk_{23456} C_{k23456}J_{k23456}\delta^{k23}_{456} \right),
\end{equation}
where $J_{k23456}\delta^{k23}_{456} = \langle c^*_kc_2^*c_3^*c_4c_5c_6 \rangle$.

Assuming a Gaussian wave field, one can take $J_{k23456}$ to the zeroth order $J^{(0)}_{k23456}$, which is simplified via Gaussian statistics to a product of three pair correlators,
\begin{eqnarray}\label{eq:rpa}
J^{(0)}_{k23456}&=&n_2n_3n_4\big[\delta^k_4\left(\delta^2_5\delta^3_6+\delta^2_6\delta^3_5\right)\nonumber\\
&+&\delta^k_5\left(\delta^2_4\delta^3_6+\delta^2_6\delta^3_4\right)+\delta^k_6\left(\delta^2_4\delta^3_5+\delta^2_5\delta^3_4\right)\big].
\end{eqnarray}
However, due to the symmetry of $C_{k23456}$ this makes the right hand side of kinetic equation (\ref{eq:ke}) zero.  To find a nontrivial answer we need to obtain a first
order addition $J^{(1)}_{k23456}$ to $J_{k23456}$. To calculate $J^{(1)}_{k23456}$ one takes the time derivative of $J_{k23456}$, using the equation of motion (\ref{eq:cdot}) of the canonical variable $c_k$, and insert the zeroth order approximation for the tenth correlation function (this is similar to equation (\ref{eq:rpa}), but a product of five pair correlators involving ten wavevectors). $J^{(1)}_{k23456}$ can then be written as

\begin{equation}\label{eq:integratedJ}
J^{(1)}_{k23456}=Be^{i\Delta\omega t}+\frac{A_{k23456}}{\Delta\omega},
\end{equation}
where $\Delta\omega=\omega_k+\omega_2+\omega_3-\omega_4-\omega_5-\omega_6$ and \\ $A_{k23456}= 3C^*_{k23456}n_k n_2 n_3 n_4 n_5 n_6 \left[{1 \over n_k}+{1 \over n_2}+{1 \over n_3}-{1 \over n_4}-{1 \over n_5}-{1 \over n_6} \right]$.  The first term of (\ref{eq:integratedJ}) is a fast oscillating function, its contribution to the integral (\ref{eq:n_k}) decreases with $z$ and is negligible at $z$ larger than $1/\omega_k$, and as a result we will ignore the contribution arising from this term. The second term is substituted back into equation (\ref{eq:n_k}), the relation $Im(\Delta\omega)\sim -\pi\delta(\Delta\omega)$ is applied because of integration around the pole, and the kinetic equation is derived,

\begin{eqnarray}
\dot{n}_k &=& 18 \pi \int dk_{23456} \; |C_{k23456}|^2 \;
\delta^{k23}_{456} \; \delta(\omega^{k23}_{456}) \; f_{k23456},
\label{eq:ke}
\end{eqnarray}
where we have introduced
$f_{k23456}=n_k n_2 n_3 n_4 n_5 n_6 \left[{1 \over n_k}+{1 \over n_2}
+{1 \over n_3}-{1 \over n_4}-{1 \over n_5}-{1 \over n_6} \right]$ and $\delta(\omega^{k23}_{456})=\delta(\omega_k+\omega_2+\omega_3-\omega_4-\omega_5-\omega_6)$.

A simple dimensional analysis of (\ref{eq:ke}) gives
\begin{equation}
\dot{n}_k \sim k^{14} n_k^5,
\label{eq:dim}
\end{equation}
which is the same form obtained from the full BSE \cite{KS04}.

In wave turbulence theory, one is concerned with non-equilibrium steady state solutions of the kinetic equation (\ref{eq:ke}).  These solutions, that rely on a constant (non-zero) flux in some inertial range are known as Kolmogorov-Zakharov (KZ) solutions.  In addition, the kinetic equation (\ref{eq:ke}) contains additional solutions that correspond to the thermodynamical equipartition of energy and wave action.  These equilibrium solutions stem from the limiting cases of the more generalised Rayleigh-Jeans distribution

\begin{equation}\label{eq:rj}
n_k = \frac{T}{\omega_k + \mu },
\end{equation}
where $T$ is the temperature of the system and $\mu$ is a chemical potential.

To find the KZ solutions, one can apply a dimensional argument on both energy and wave action fluxes.
The energy flux at wavenumber $k$ is defined as
$\Pi^{(H)}_k = \int dk' \dot{n}_{k'} \omega_{k'}$ which, using
(\ref{eq:dim}), becomes $\Pi^{(H)}_k \sim k^{17} n_{k}^5$.
By requiring the existence of a range of scales in which the energy flux is
$k$-independent leads to the spectrum
\begin{equation}
n_k \sim k^{-17/5},
\label{eq:spdir}
\end{equation}
which, again,  is the same form obtained from the full BSE \cite{KS04}.

A similar argument can be applied to the wave action (\ref{eq:n})
whose flux is $\Pi^{(N)}_k = \int dk' \dot{n}_{k'} \sim k^{15} n_{k}^5$.
Therefore a scale independent flux of wave action requires a spectrum  \cite{lebedev,N06}
\begin{equation}
n_k \sim k^{-3},
\label{eq:spinv}
\end{equation}
A word of caution is due about both spectra (\ref{eq:spdir}) and (\ref{eq:spinv})
because the dimensional analysis does not actually guarantee that they
are true solutions of the kinetic equation. To check if these spectra
are real solutions (and therefore physically relevant)  one has to prove their
{\em locality} i.e. convergence of the kinetic equation integral on these
spectra.  This has not been done yet, neither for the full BSE nor for the truncated-LIA model,
and this is especially worrying since the spectrum (\ref{eq:spdir}) has already been
accepted by sizable part of the quantum turbulence community and has been used
in further theoretical constructions.
We announce that there is a work in progress to check locality of spectra (\ref{eq:spdir}) and (\ref{eq:spinv})
in both the BSE and truncated-LIA settings.
However, as we will see later, at least for the truncated-LIA these spectra are observed
numerically, so we will tentatively assume that they are true and relevant solution.

The two spectra (\ref{eq:spdir}) and (\ref{eq:spinv}) occur in different
scale ranges and
the two cascades develop in opposite directions, as in the case
of two-dimensional turbulence \cite{KM80}. Among the two conserved
quantities, the largest contribution to energy comes from smaller scales
than those that contribute to wave action (because the former
contains the field derivatives). Therefore, according to
the Fj{\o}rtoft argument \cite{Fjortoft53}, we expect to have
a {\it direct cascade} of energy with a $k^{-17/5}$ spectrum flowing to large $k$ and
an {\it inverse cascade} of wave action with spectrum $k^{-3}$ flowing to small $k$.

\section{Numerical Results for Truncated-LIA}
In the following we will consider numerical simulations of the
system (\ref{eq:expansion}) under the conditions in which
a stationary turbulent cascade develops. Energy and wave action are
injected in the vortex filament by a white-in-time external forcing
$\phi(z,t)$ acting on a narrow band of wavenumbers around a given $k_f$.
In order to have a stationary cascade, we need additional terms which
remove $H$ and $N$ at small and large scales. The equation of motion
obtained from (\ref{eq:expansion}) is therefore modified as
\begin{equation}
\dot{w} = {i \over 2} \left[ w' \left( 1 - {1 \over 2} |w'|^2 \right)
\right]' - (-1)^p \nu \nabla^{2p} w - \alpha w + \phi.
\label{eq:liatot}
\end{equation}
In (\ref{eq:liatot})
the small scale dissipative term (with $p > 1$) physically
represents the radiation of phonons (at a rate proportional to
$\nu$) and the large scale damping term can be interpreted
as the friction induced by normal fluid at a rate $\alpha$.

Assuming the spectra (\ref{eq:spdir}) and (\ref{eq:spinv}),
a simple dimensional analysis gives the IR and UV cutoff induced
by the dissipative terms. The direct cascade is removed
at a scale $k_{\nu} \sim \nu^{-5/(10p-2)}$ while the inverse cascade
is stopped at $k_{\alpha} \sim \alpha^{1/2}$. Therefore, in an
idealized realization of infinite resolution one would obtain a
double cascade by keeping $k_f=O(1)$ and letting $\nu,\alpha \to 0$.
In order to have an extended inertial range, in finite resolution
numerical simulations we will restrict ourselves to resolve a single cascade
at a time by putting either $k_f \simeq k_{\alpha}$ or $k_f \simeq k_{\nu}$ for
the direct and inverse cascades respectively.

We have developed a numerical code which integrates the
equation of motion (\ref{eq:liatot}) by means of a pseudospectral
method for a periodic vortex filament of length $2 \pi$ with a
resolution of $M$ points.
The linear and dissipative terms are integrated explicitly
while the nonlinear term is solved by a second-order Runge-Kutta
time scheme. The vortex filament is initially a straight line
($w(z,t=0)=0$) and long time integration is performed until a
stationary regime (indicated by the values of $H$ and $N$)
is reached.
The ratio between the two terms in the series (\ref{eq:expansion})
is $H_1/H_2 \simeq 20$, confirming {\it a posteriori} the validity
of the perturbative series (\ref{eq:expansion}) and the condition of the small amplitude perturbation $\epsilon \ll 1$ in the derivation of equation (\ref{eq:h2d}).

The first set of simulations is devoted to the study of the
direct cascade.
Energy fluctuations are injected at a forcing wavenumber
$k_f \simeq 2$ and the friction coefficient $\alpha$ is set in order
to have $k_{\alpha} \simeq k_f$.
Energy is removed at small scales by hyperviscosity of order $p=4$ which
restricts the range of dissipation to the wavenumber in a close vicinity of $k_{max}$.

\begin{figure}[htp]
\begin{center}
\includegraphics[scale=0.7]{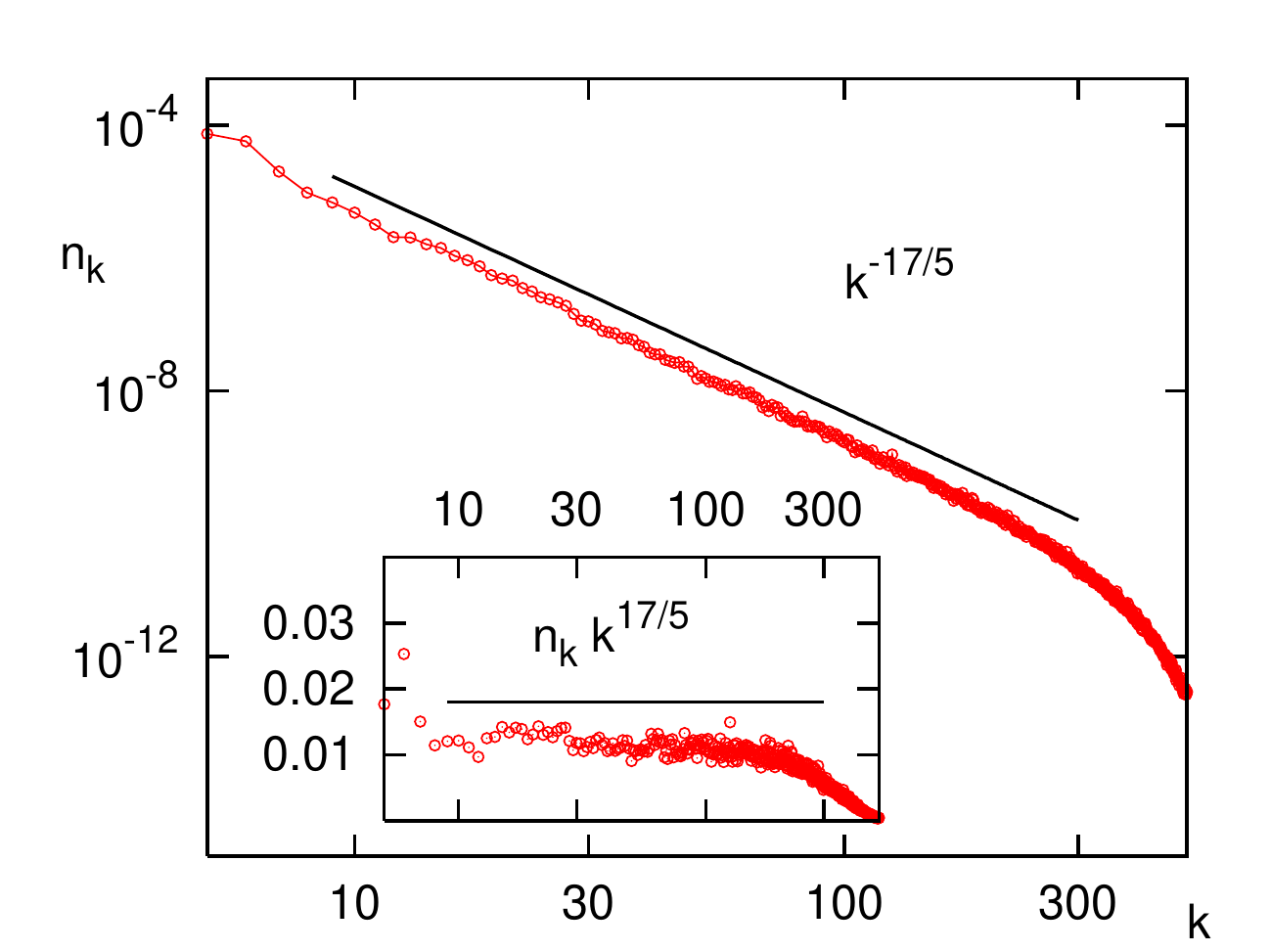}
\caption{Wavenumber spectrum $n_k$ for a simulation of the direct
cascade in stationary conditions at resolution $M=2048$. Forcing
is restricted to a range of wavenumbers $1 \le k_f \le 3$ and
dissipation by phonon emission is modeled with hyperviscosity of order $p=4$.
The straight line represents the kinetic equation prediction
$n_k \simeq k^{-17/5}$. The inset shows the spectrum compensated
with the theoretical prediction.}
\label{fig1}
\end{center}
\end{figure}

In Figure~\ref{fig1} we plot the wave action spectrum for the
direct cascade simulation, averaged over time in stationary conditions.
A well developed power law spectrum very close
to prediction (\ref{eq:spdir}) is observed over more than one
decade (see inset). This spectrum confirms the existence of
non-trivial dynamics with six-wave processes for the truncated
Hamiltonian (\ref{eq:expansion}).


The direct cascade of the full Biot-Savart Hamiltonian (\ref{eq:h2d}) was discussed by Kozik and Svistunov who gave the dimensional prediction (\ref{eq:spdir}) \cite{KS04}, who later performed a numerical simulation of the nonlocal BSE to confirm the scaling \cite{KS05}.

\begin{figure}[htp]
\begin{center}
\includegraphics[scale=0.7]{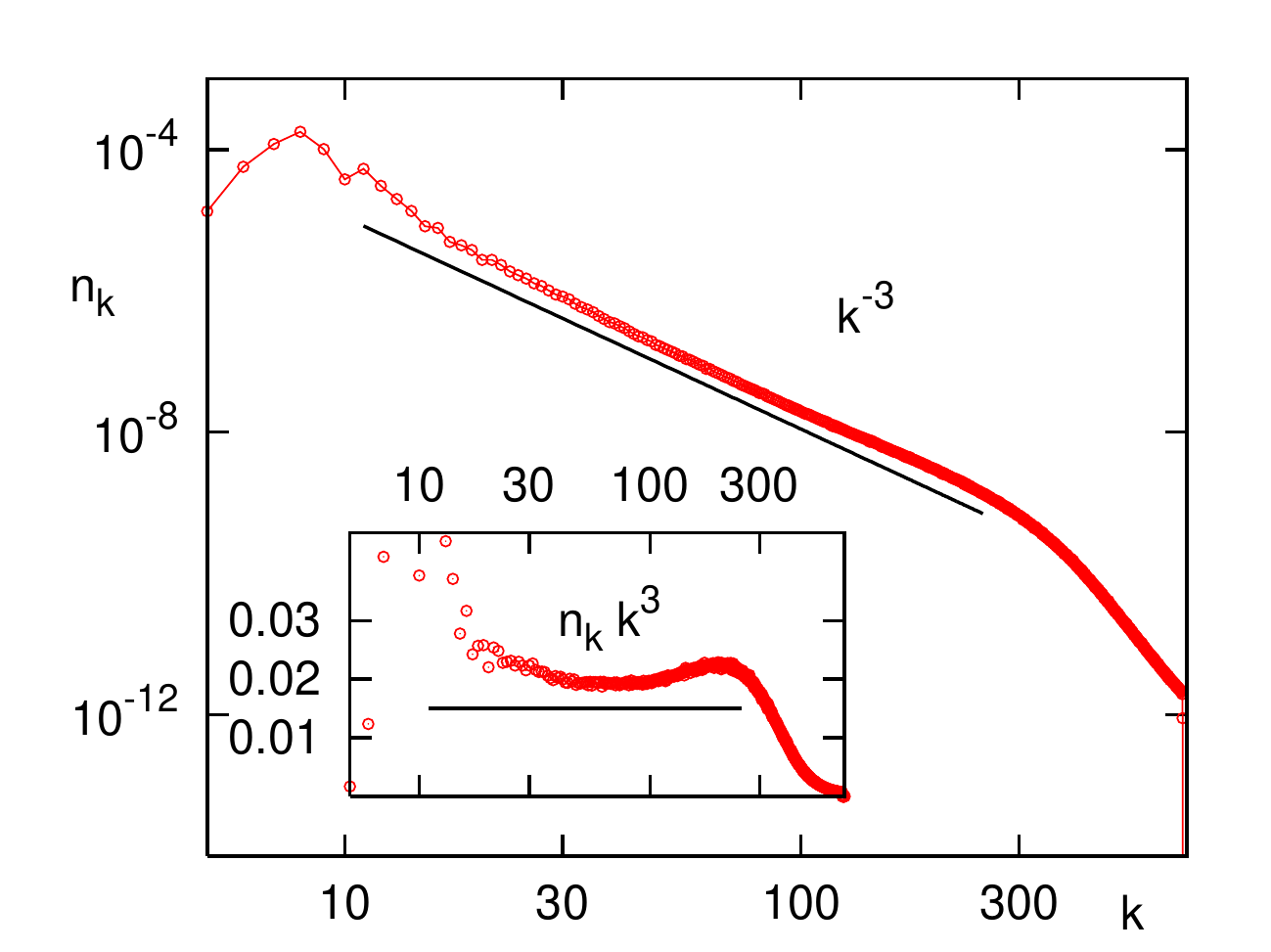}
\caption{Wavenumber spectrum $n_k$ for a simulation of the inverse
cascade in stationary conditions at resolution $M=2048$. Forcing
is restricted to a range of wavenumbers around $k_f=300$ and
dissipation by phonon emission is modelled with hyperviscosity of order $p=4$.
The straight line represents the kinetic equation prediction
$n_k \simeq k^{-3}$. The inset shows the spectrum compensated
with the theoretical prediction.}
\label{fig2}
\end{center}
\end{figure}

We now turn to the simulation for the inverse cascade regime.
To obtain an inverse cascade, forcing is concentrated at small scales,
here $k_f=300$. In order to avoid finite size effects and accumulation
at the largest scale \cite{SY94}, the friction coefficient is chosen in such a
way that wave action is removed at a scale $k_{\alpha}\simeq 10$.
Figure~\ref{fig2} shows the spectrum for
this inverse cascade in stationary conditions.
In the compensated plot, a small deviation from the power-law scaling
at small wavenumbers  is observed, probably due to the presence of condensation (``inverse bottleneck") .
Nevertheless, a clear scaling compatible with the dimensional analysis
of the kinetic equation is observed over about a decade.

\section{Differential Approximation Model}

Differential Approximation Models (DAMs) have proved to be a very useful tool in the analysis of fluid
dynamical and wave turbulence \cite{L68,H85,I85,ZP99,c04,L04,N06,L06}.  These equations are constructed using a differential closure, such that the main scalings of the original closure (kinetic equation in our case) are preserved.  In addition, there exist a family of simpler or reduced DAMs for which rigourous analysis can be performed upon their solutons. These appears to be quite helpful when the full details about the  dynamics are not needed.  Moreover, due to the DAMs simplicity, one can add physically relevant forcing and dissipative terms to the models.

For the Kelvin wave spectra (\ref{eq:spdir}) and (\ref{eq:spinv}), including the thermodynamical equipartition solutions (\ref{eq:rj}), the corresponding DAM is \cite{N06}
\begin{equation}\label{completeDAM}
\frac{\partial n_\omega}{\partial t} = \frac{C}{\kappa^{10}}\omega^{1/2}\frac{\partial^2}{\partial \omega^2}\left(n^6_\omega\omega^{21/2}\frac{\partial^2}{\partial \omega ^2}\frac{1}{n_\omega}\right),
\end{equation}
where $\kappa$ is the vortex line circulation, $C$ is a dimensionless constant and $\omega=\omega(k)=\kappa k^2 /{4\pi}$ is the Kelvin wave frequency.  Notice, that the DAM is written in terms of frequency $\omega$, rather than the wavenumber $k$, as in the case of the kinetic equation (\ref{eq:ke}).

Equation (\ref{completeDAM}) preserves the energy
\begin{equation}
E = \int \omega^{1/2}n_\omega d \omega,
\end{equation}
and wave action
\begin{equation}
N = \int \omega^{-1/2}n_\omega d \omega.
\end{equation}

The forcing of Kelvin waves on quantized vortices arises from sharp cusps produced by vortex reconnections \cite{N06}.  These reconnections can be interpreted in the DAM by the addition of the term \cite{N06}
\begin{equation}
\left[\frac{\partial n_\omega}{\partial t}\right]_{forcing}=\lambda \omega^{-2}.
\end{equation}
The transfer of energy flux in the Kelvin wave turbulence proceeds towards high wavenumbers until, the Kelvin wave frequencies become large enough to excite phonons in the fluid and thus dissipate the energy of the Kelvin waves into the surrounding fluid.  One can introduce sound dissipation derived from the theory of Lighthill in classical hydrodynamical turbulence \cite{L52}, and in the context of quantum turbulence has been done in \cite{KS05-2}.  This corresponds to the addition of the following term to the DAM \cite{N06}
\begin{equation}\label{}
\left[\frac{\partial n_\omega}{\partial t}\right]_{radiation}=-\nu\omega^{5}n_\omega^2.
\end{equation}
(This expression is slightly corrected with respect to the relation $\sim \omega^{9/5}n_\omega^2$ of the one  introduced dimensionally in \cite{N06}
to make it consistent with the more rigorous analysis of \cite{KS05-2}).

Finally, one can add an addition term that describes the effect of friction with the normal fluid component  as follows,
\cite{L04,L06}
\begin{equation}
\left[\frac{\partial n_\omega}{\partial t}\right]_{friction}=-\alpha\omega n_\omega.
\end{equation}

Thus, one may write a generalized DAM as
\begin{equation}\label{general}
\frac{\partial n_\omega}{\partial t}=F(n_\omega, \omega)+\lambda\omega^{-2}-\nu\omega^{5}n_\omega^2-\alpha\omega n_\omega.
\end{equation}
where the nonlinear function $F(n_\omega, \omega)$ is the interaction term, which in the case of the complete DAM is the RHS of equation (\ref{completeDAM}).

Other (reduced) DAMs include either, solutions for the direct and inverse cascade and no (thermo) equipartition solutions as in
\begin{equation}\label{DAMinv}
F(n_\omega,\omega)= \omega^{-1/2} \partial_\omega \big(n_\omega^4 \omega^8 \partial_\omega(\omega^{3/2} n_\omega) \big),
\end{equation}
or just the direct energy cascade and the corresponding energy thermo solution (and no inverse cascade solutions):
\begin{equation}\label{DAMwarm}
F(n_\omega,\omega)= \omega^{-1/2} \partial_\omega \big(n_\omega^4 \omega^{17/2} \partial_\omega(\omega n_\omega) \big).
\end{equation}

These have the advantage of only containing second order derivatives, and as such one may find analytical solutions for the steady state dynamics.  For example, for
model (\ref{DAMinv}) we can ask the question, how does the vortex reconnection forcing build the energy flux in frequency space?
 For this, we leave the nonlinear transfer and the reconnection forcing terms and drop the dissipation term, and find a solution for the energy flux $\epsilon_\omega$:
\begin{equation}\label{e-flux}
      \epsilon_\omega = \epsilon_0 -\lambda \omega^{-1/2}
\end{equation}
and the wave action density $n_\omega$:
\begin{equation}\label{sol}
      n_\omega = 10^{1/5} \omega^{-3/2} \big(\epsilon_0 \omega^{-1} -\frac{2}{3} \lambda \omega^{-3/2} -\eta_0 \big)^{1/5}
\end{equation}
where $\epsilon_0$ and $\eta_0$ are the asymptotic values of the energy and wave action fluxes respectively.

For equation (\ref{DAMwarm}), without any forcing or dissipation terms,
we find that the corresponding analytical steady state solution is
\begin{equation}\label{bottle1}
n_\omega = A\omega^{-17/10}+B\omega^{-1},
\end{equation}
which is a ``warm cascade" solution, i.e. a direct energy cascade gradually
transitioning into a thermalized ``bottleneck" over a range of scales
\cite{c04,N06}.

\section{Numerical Results for DAMs}

We performed numerical simulations of the  DAMs (\ref{completeDAM}), (\ref{DAMinv}) and (\ref{DAMwarm}), using a second order finite difference method.  We set the resolution at $M=1024$ points, while the phonon radiation dissipation term acts in the range $\omega > \omega_\nu = 800$.
The parameters $\lambda$, $\nu$ and the estimated asymptotic values of the fluxes $\epsilon_0$, $\eta_0$ are listed in Table 1.
The factor $C/\kappa^{10}$ of equation (\ref{completeDAM}) has been fixed to unity.

\begin{table}[htbp]
\begin{center}
\begin{tabular}{|l|c|c|} \hline
                  & Reduced DAM                        & Complete DAM \\ \hline
Resolution        & $M=1024$                           & $M=1024$ \\
Forcing amplitude & $\lambda=2\times10^{-9}$            & $\lambda=1\times10^{-9}$ \\
Viscosity         & $\nu=2\times10^{-12}$               & $\nu=1\times10^{-12}$ \\
Energy flux       & $\epsilon_0\simeq1.66\times10^{-9}$ & $\epsilon_0\simeq0.66\times10^{-9}$ \\
Waveaction flux   & $\eta_0\simeq1.8\times10^{-12}$     & $\eta_0\simeq0.75\times10^{-12}$ \\ \hline
\end{tabular}
\caption{Parameters and fluxes of the simulations for DAM models with $\nu \ne 0$}
\label{table}
\end{center}
\end{table}

\begin{figure}[htp]
\begin{center}
\includegraphics[scale=0.5]{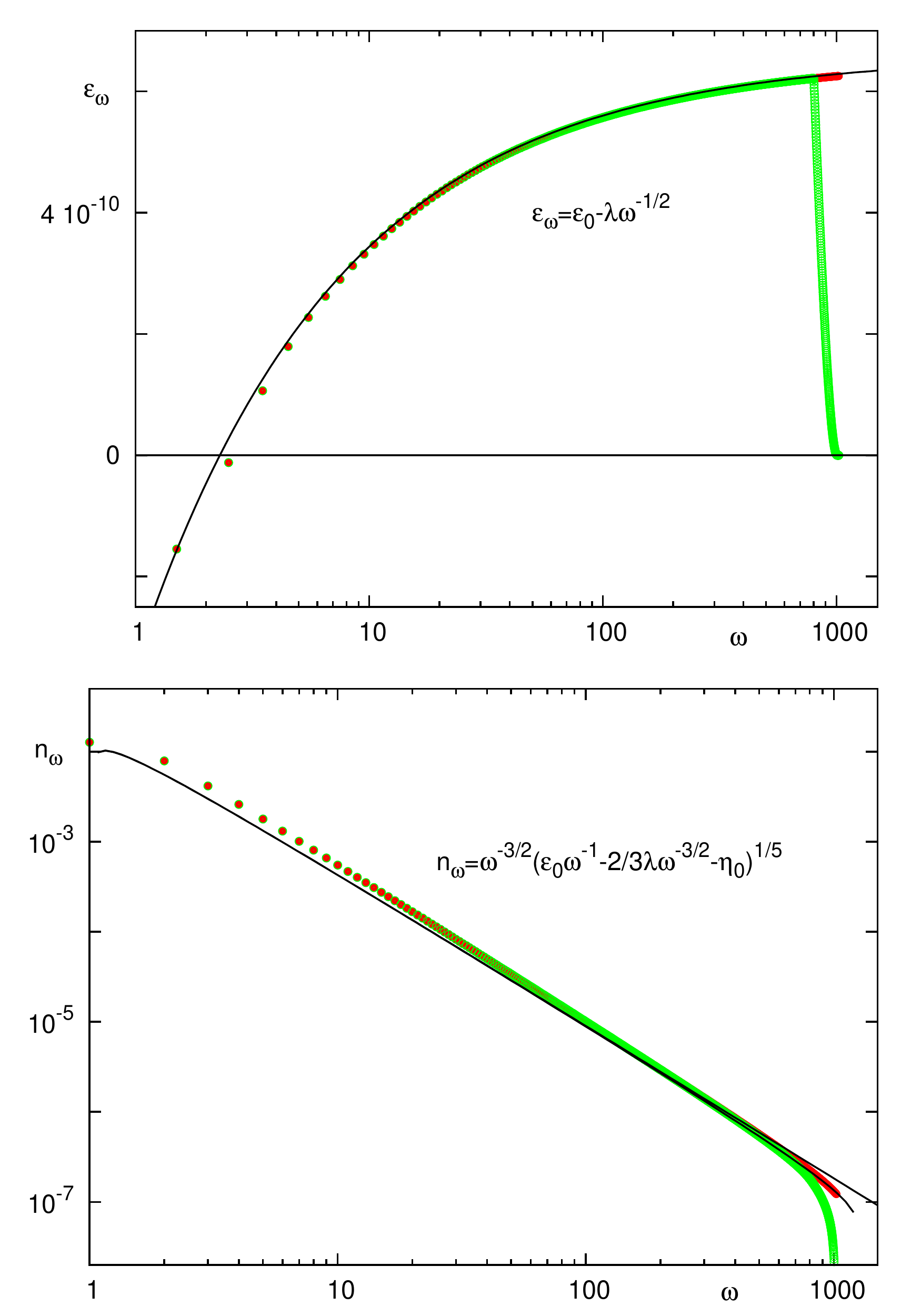}
\caption{Complete DAM: the energy flux $\varepsilon$ (top) and the spectrum (bottom),  compared with the analytical predictions (\ref{e-flux}) and (\ref{sol})  respectively. In each picture  the results of the simulations with and without  phonon dissipation are shown (the phonon dissipation can be seen as an abrupt cutoff of the flux and the spectrum).
The spectrum is very close to the prediction of the reduced model (\ref{DAMinv}): $n_\omega\sim\omega^{-1.73}$ compared with  $n_\omega\sim\omega^{-1.7}$ which is expected  in the inertial range ($ 10 < \omega < 800$).}
\label{fig3}
\end{center}
\end{figure}

Initially, we simulate both the complete DAM (\ref{completeDAM}) and the reduced DAM  (\ref{DAMinv}), with forcing, without friction, but with and without the dissipative term for phonon radiation. The results for the energy flux and the spectrum for the complete DAM are shown in Figure~\ref{fig3} (results for the reduced DAM are nearly identical and, therefore, are not shown).
 The top panel in Figure~\ref{fig3} shows the energy flux. We see a good  agreement with the analytical prediction (\ref{e-flux}) over a large
  intermediate range of scales.
  In the case without the phonon dissipation,
  the numerical result for the energy flux follows perfectly the analytical prediction at high frequencies.
  In the case with the phonon dissipation, the agreement with the analytical prediction (\ref{e-flux}) is good in a long range
  up to very high frequencies, where the phonon dissipation suddenly kicks in. Such a sudden onset of dissipation is due to
  the abrupt growth of the phonon radiation term as a function of the frequency.
   The wave action spectra are shown in the bottom panel of Figure~\ref{fig3}. We
   see a good agreement with the analytical solution (\ref{sol}) where $\eta_0$ is taken to be zero (flux of
   wave action $\eta_0$ could only be generated by an extra forcing at the high-frequency end, which is absent in our case).
    We observe only a slightly steeper spectrum of $n_\omega\sim \omega^{-1.73}$ compared with the analytical prediction of the direct energy cascade $n_\omega\sim \omega^{-17/10}$.
 This agreement is remarkable because the solution (\ref{DAMinv}) is strictly valid only for the reduced and not the complete DAM.
 This shows that the reduced DAM does pretty well in predicting the behaviour of the more complete nonlinear model.
    Naturally, in the case with the phonon dissipation, we see a deviation from the analytical solution at very small scales (a rather sharp cut-off).

\begin{figure}[htp]
\begin{center}
\includegraphics[scale=0.4]{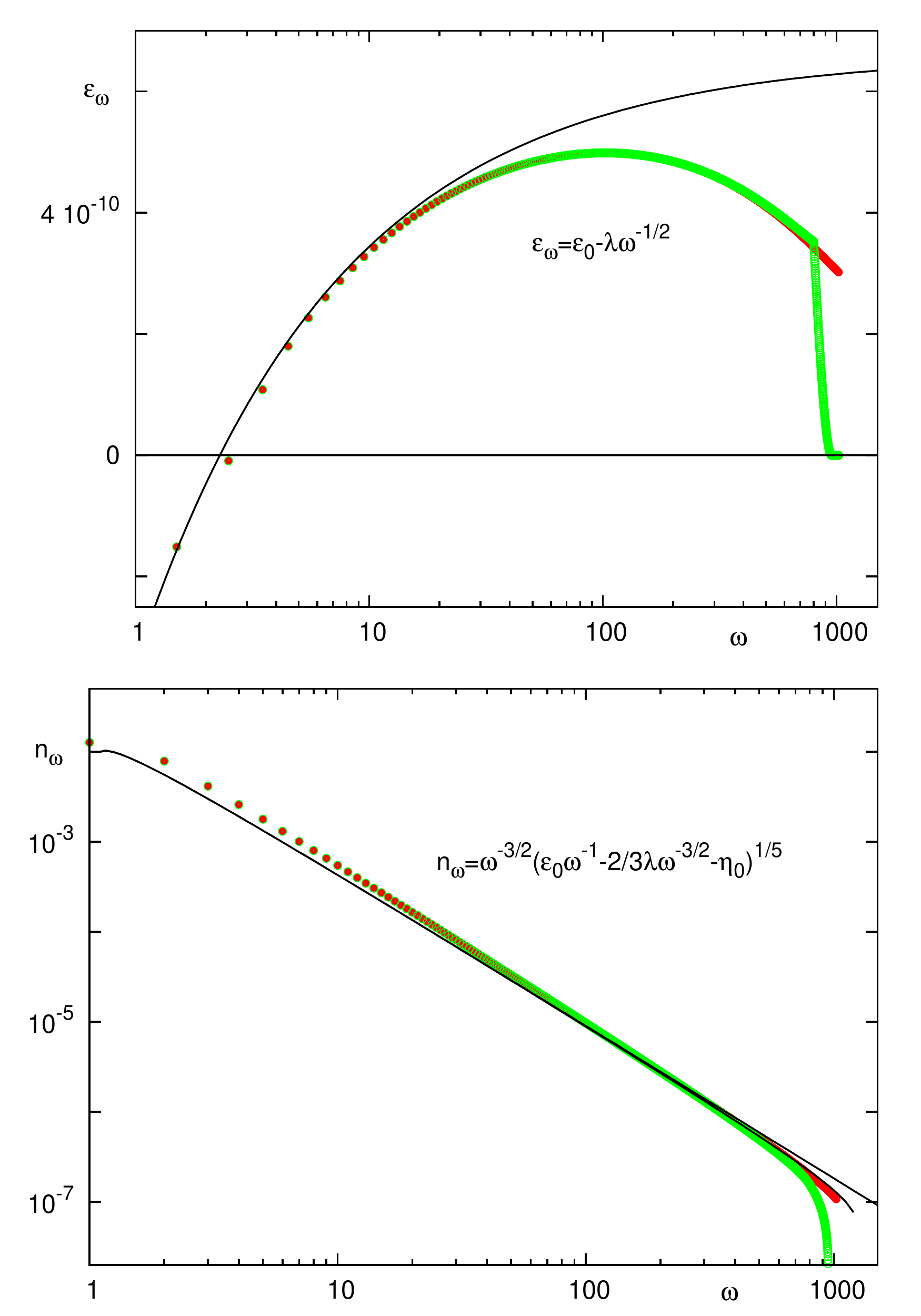}
\caption{Complete DAM with $\alpha \ne 0$ and $\nu=0$ or $\nu \ne 0$. The energy flux is reduced at high $\omega$'s,
but the spectrum slope remains as in the non-dissipative case for a longer frequency range.}
\label{fig4}
\end{center}
\end{figure}

In Figure~\ref{fig4}
we show the effect of switching on the friction term ($\alpha \neq 0$) for the complete DAM (again, results for the reduced DAM are very similar and are not shown).
  We see
 that the presence of the friction has the effect of reducing the energy flux in a large region from frequencies of around $\omega \approx 10$ upwards.  Although the flux is  reduced, we see  that the spectrum has only a slight deviation from the predicted slope (\ref{sol}).

\begin{figure}[htp]
\begin{center}
\includegraphics[scale=0.5]{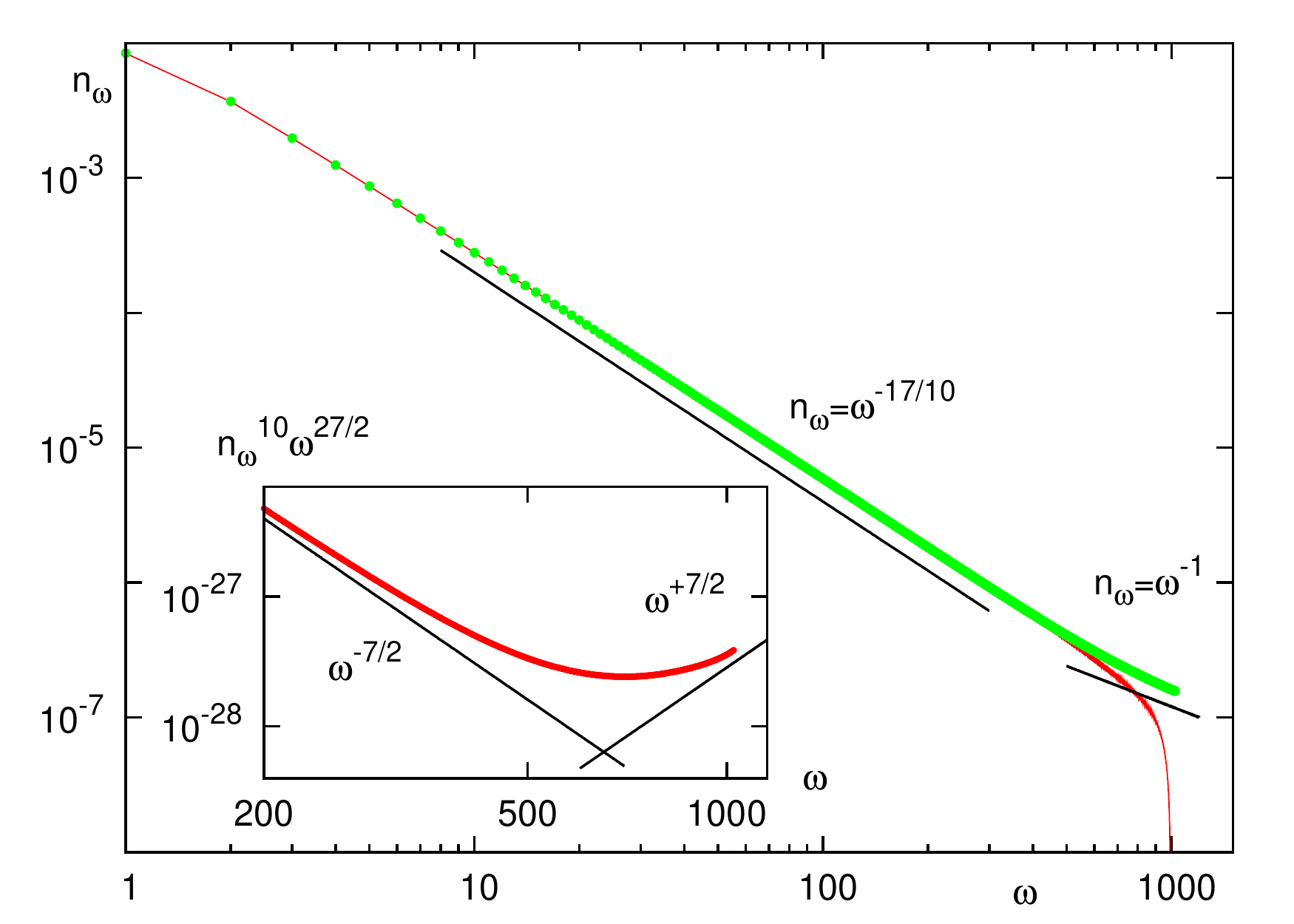}
\caption{Bottleneck effect for the second order model \cite{N06}. The spectrum ($n(\omega)$ w.r.t. $\omega$) compared with the two different power law predictions in the two ranges and the local slope analysis in the inset ($a$ w.r.t. $\omega$) which points out the transition from the $n_\omega \sim \omega^{-1.7}$ behaviour to the $n_\omega \sim \omega^{-1}$ one. In the picture below, we show a zoom of the region where the bottleneck effect appears in a sort of compensated spectrum, $n_\omega^{10} \omega^{27/2}$ on the y-axis and $\omega$ on the x-axis.}
\label{fig5}
\end{center}
\end{figure}

Finally we consider another DAM model, (\ref{DAMwarm}), which contains only the direct and the thermodynamical bottleneck (or warm cascade) of energy.  We force the system as usual, however, the simulation is composed of two phases: the first is to get a direct cascade steady state, then lowering the viscosity $\nu$ (i.e. dissipation at small scales) and then turning on the reflecting energy flux boundary condition at the smallest scale $\omega=\omega_{MAX}=M$, after which the system evolves to a secondary steady state (\ref{bottle1}).
In Figure~\ref{fig5} we see a clear transition from the KZ solution towards the thermalized spectrum.  The bottom panel is a zoomed in section of the crossover using a compensated spectrum, so that one can clearly make the distinction between the two power laws.  We see a good agreement with the predicted power law behaviour of the KZ solution and of the thermalized solution.

\newpage

\section{Conclusions}
In summary, we have introduced and studied various reduced models for Kelvin wave turbulence. Firstly, we introduced a truncated-LIA model for Kelvin wave turbulence, which we have shown to exhibit the same scalings and dynamical features present in the conventional BSE.
We have used this model for numerical simulations of the direct and the
inverse cascades and found spectra which are in very good agreement with the
predictions of the WT theory. Secondly, we discussed differential approximation models, and introduced three such models to describe various settings of Kelvin wave turbulence, such as the direct energy cascade generated by vortex reconnections and dissipated via phonon radiation or/and the mutual friction with the normal
 liquid, as well as  the bottleneck effect when the energy flux is reflected from the smallest scale.
   We performed numerical simulations of these cases, which showed good agreements with the predicted analytical solutions.

\newpage

\section{Appendix A - Interaction Coefficients in Biot-Savart model}

In this Appendix we review and extend the work of Kozik and Svistunov on the Kelvin wave cascade (KS$04$) \cite{KS04}.  They considered the full Biot-Savart Hamiltonian (\ref{eq:h2d}) in $2$D, and simplified the denominator by Taylor expansion.  The criterion for Kelvin-wave turbulence is that the wave amplitude is small compared to wavelength, this is formulated as:
\begin{equation}
\epsilon(z_1,z_2)=\frac{|w(z_1)-w(z_2)|}{|z_1-z_2|}\ll 1.
\end{equation}

KS$04$ find the Biot-Savart Hamiltonian (\ref{eq:h2d}) expanded in powers of $\epsilon$ ($H={H}_0+{H}_1+{H}_2+{H}_3$, here ${H}_0$ is just a number and is ignored) is represented as:

\begin{eqnarray}\label{eq:KS04Ham}
{H}_1&=&\frac{\kappa}{8\pi}\int \frac{d z_1 d z_2}{|z_1-z_2|}\left[2Re\left(w^{'*}(z_1)w^{'}(z_2)\right)-\epsilon^2\right],\nonumber\\
{H}_2&=&\frac{\kappa}{32\pi}\int \frac{d z_1 d z_2}{|z_1-z_2|}\left[3\epsilon^4-4\epsilon^2Re\left(w^{'*}(z_1)w^{'}(z_2)\right)\right],\\
{H}_3&=&\frac{\kappa}{64\pi}\int \frac{d z_1 d z_2}{|z_1-z_2|}\left[6\epsilon^4Re\left(w^{'*}(z_1)w^{'}(z_2)\right)-5\epsilon^6\right]\nonumber.
\end{eqnarray}

One would like to deal with the wave-interaction Hamiltonian (\ref{eq:KS04Ham}) in Fourier space by introducing the wave amplitude variable $w_k$.
Using the Fourier representation for variables  $w(z_1)=\int dk w_k e^{ikz_1}$ and $w(z_2)=\int dk w_k e^{ikz_2}$ in equations (\ref{eq:KS04Ham}), one introduces more integration variables, the wavenumbers. Moreover, invoking a cutoff at $a<|z_1-z_2|$ because of the singularity present in the Biot-Savart Hamiltonian (\ref{eq:h2d}) as $|z_1-z_2|\rightarrow 0$, KS$04$ derived the coefficients of $H_1$, $H_2$ and $H_3$ in terms of cosines in Fourier space \cite{KS04}. Once the equations (\ref{eq:KS04Ham}) are written in terms of wave amplitudes, one can define variables $z_-$ and $z_+$ as variables $z_-=|z_1-z_2|$ which ranges from $a$ to $\infty$ and $z_+=z_1+z_2$ ranging from $-\infty$ to $\infty$. One can then decomposes all variables of type $z_1$ and $z_2$ into variables $z_+$ and $z_-$. The cosine functions arise due to the collection of exponentials with powers in variable $z_-$.  Subsequently, the remaining exponentials with powers of variable $z_+$ can be integrated out w.r.t. $z_+$, yielding the corresponding delta function for the conservation of wavenumbers.  The explicit formulae of the four-wave $W_{1234}$, and the six-wave $T_{123456}$ interaction coefficients derived by Kozik and Svistunov can be written as:

\begin{eqnarray}
\omega_k &=& \frac{\kappa}{2\pi}\left[A-B\right],\label{KSOmega}\\
W_{1234}&=&\frac{\kappa}{16\pi}\left[6D-E\right],\label{KSW}\\
T_{123456}&=&\frac{\kappa}{16\pi}\left[3P-5Q\right]\label{KST},
\end{eqnarray}
where $A$, $B$, $D$, $E$, $P$ and $Q$ are integrals of cosines:

\begin{eqnarray}
A &=& \int_a^\infty \frac{dz_-}{z_-}k^2C^k,\label{eq:A}\\
B &=& \int_a^\infty \frac{dz_-}{z^3_-}\left[1-C^k\right],\label{eq:B}\\
D &=& \int_a^\infty \frac{dz_-}{z_-^5}\left[1-C_1-C_2-C^3-C^4+C^3_2+C^{43}+C^4_2\right],\label{eq:D}\\
E &=& \int_a^\infty \frac{dz_-}{z_-^3}[k_1k_4\left(C^4+C_1-C^{43}-C^4_2\right)+k_1k_3\left(C^3+C_1-C^{43}-C^3_2\right)+\nonumber\\
&& +k_3k_2\left(C^3+C_2-C^{43}-C^3_1\right)+k_4k_2\left(C^4+C_2-C^{43}-C^3_2\right)],\label{eq:E}\\
P &=& \int_a^\infty\frac{dz_-}{z_-^5}k_6k_2[C_2-C^5_2-C_{23}+C^5_{23}-C^4_2+C^{45}_2+C^4_{23}-C^6_1+C^6-C^{56}-\nonumber\\
&&-C^6_3+C^{56}_3-C^{46}+C^{456}+C^{46}_3-C_{12}],\label{eq:P}\\
Q &=& \int_a^\infty\frac{dz_-}{z_-^7}[1-C^4-C_1+C^4_1-C^6+C^{46}+C^6_1-C^{46}_1-C^5+C^{45}+C^5_1-C^{45}_1+\nonumber\\
&&+C^{65}-C^{456}-C^{56}_1+C_{23}-C_3+C^4_3+C_{13}-C^4_{13}+C^6_3-C^{46}_3-C^6_{13}+C^5_2+\nonumber \\
&&+C^5_3-C^{45}_3-C^5_{13}+C^6_2-C^{56}_3+C_{12}+C^4_2-C_2]\label{eq:Q},
\end{eqnarray}
where the variable, $z_- = |z_1-z_2|$ and the expressions $C$, are cosine functions such that $C_1 = \cos(k_1z_-)$, $C^4_1 = \cos((k_4-k_1)z_-)$, $C^{45}_1=\cos((k_4+k_5-k_1)z_-)$, $C^{45}_{12}=\cos((k_4+k_5-k_1-k_2)z_-)$ and so on.

We integrate the Fourier representations of $H_1$, $H_2$ and $H_3$, namely equations (\ref{eq:A}), (\ref{eq:B}), (\ref{eq:D}), (\ref{eq:E}), (\ref{eq:P}) and (\ref{eq:Q}) using integration by parts, and apply the following cosine identity \cite{GR80},

\begin{eqnarray}
\int_a^\infty\frac{\cos(z)}{z}dz &=& -\gamma-\ln(a)-\int_0^a\frac{\cos(z)-1}{z}dz\nonumber\\
&=& -\gamma-\ln(a)-\sum_{k=1}^\infty\frac{\left(-a^2\right)^k}{2k\left(2k\right)!}\nonumber\\
&=& -\gamma-\ln(a)+\mathcal{O}(a^2).
\end{eqnarray}
Neglecting terms of order $\sim a^2$ and higher we calculate the frequency and interaction coefficients of equations (\ref{eq:KS04Ham})
\begin{eqnarray}
{H}_1 &=&\int \omega_k |w_k|^2 d k,\\
{H}_2 &=&\int d k_{1234}W_{1234}w^*_1w^*_2w_3w_4\delta^{12}_{34} ,\\
{H}_3 &=&\int d k_{123456}C_{123456}w^*_1w^*_2w^*_3w_4w_5w_6\delta^{123}_{456},
\end{eqnarray}
where
\begin{eqnarray}
\omega_k &=& \frac{\kappa}{4\pi}k^2\left[\ln\left(\frac{1}{k_{\mathrm{eff}}a}\right)-\gamma -\frac{3}{2}\right]- \frac{\kappa k^2}{4\pi}\ln\left(\frac{k}{k_{\mathrm{eff}}}\right),\label{eq:BSomega}\\
W_{1234} &=& \frac{\kappa}{64\pi}k_1k_2k_3k_4\left[1+4\gamma-4\ln\left(\frac{1}{k_{\mathrm{eff}}a}\right)\right]+F_{1234},\label{eq:BSW}\\
C_{123456}&=& \frac{\kappa}{128\pi}k_1k_2k_3k_4k_5k_6\left[1-4\gamma+4\ln\left(\frac{1}{k_{\mathrm{eff}}a}\right)\right] + G_{123456}.\label{eq:BST}
\end{eqnarray}
We use the notation that $k_{\mathrm{eff}}$ is the mean value of wavenumbers, $\gamma=0.5772\dots$ is the Euler constant and $F_{1234}$ and $G_{123456}$ are logarithmic terms of order one that are shown below,
\begin{eqnarray}
F_{1234}&=&-\frac{\kappa}{16\pi}\Big[6\sum_{N \in I}\frac{N^4}{24}\ln\left(\frac{N}{k_{\mathrm{eff}}}\right)+\sum_{N \in J}k_ik_j\frac{N^2}{2}\ln\left(\frac{N}{k_{\mathrm{eff}}}\right)\Big],
\end{eqnarray}
\begin{eqnarray}
G_{123456}&=&-\frac{\kappa}{16\pi}\Big[3\sum_{N \in K}k_6k_2\frac{N^4}{24}\ln\left(\frac{N}{k_{\mathrm{eff}}}\right) +5\sum_{N \in L}\frac{N^6}{720}\ln\left(\frac{N}{k_{\mathrm{eff}}}\right)\Big],
\end{eqnarray}
\begin{eqnarray}
I&=&\left\{ -[_1],-[_2],-[^3],-[^4],[^3_2],[^{43}],[^4_2] \right\},\\
J&=&\Big\{ \left\{[^4], [_1], -[^{43}],-[^4_2]\right\}_{i=4,j=1},\nonumber\\
&&\left\{[^3], [_1], -[^{43}],-[^3_2]\right\}_{i=3,j=1},\nonumber\\
&&\left\{[^3], [_2], -[^{43}],-[^3_1]\right\}_{i=3,j=2},\nonumber\\
&&\left\{[^4], [_2], -[^{43}],-[^4_1]\right\}_{i=4,j=2}\Big\},\\
K&=&\Big\{ [_2],-[^5_2],-[_{23}],[^5_{23}],-[^4_2],[^{45}_2],[^4_{23}],-[^6_1],[^6],-[^{56}],-[^6_3],\nonumber \\
&&[^{56}_3],-[^{46}],[^{456}],[^{46}_3],-[_{12}]\Big\},\\
L&=&\Big\{ -[^4],-[_1],[^4_1],-[^6],[^{46}],[^6_1],-[^{46}_1],-[^5],[^{45}],[^5_1],-[^{45}_1],\nonumber\\
&&[^{65}],-[^{456}],-[^{56}_1],[_{23}],-[_3],[^4_3],[_{13}],-[^4_{13}],[^6_3],-[^{46}_3],\nonumber\\
&&-[^6_{13}],[^5_2],[^5_3],-[^{45}_3],-[^5_{13}],[^6_2],-[^{65}_3],[_{12}],[^4_2],-[_2]\Big\}.
\end{eqnarray}
The notation used for the logarithmic terms is as follows, for $N\in K = \pm[^{pq}_r]$, the corresponding term in $G_{123456}$ is
\begin{equation}
\mp\frac{3\kappa}{16\pi}k_6k_2\frac{(k_p+k_q-k_r)^4}{24}\ln\left(\frac{k_p+k_q-k_r}{k_{\mathrm{eff}}}\right).
\end{equation}

We checked numerically that after the canonical transformation of the four-wave dynamics, that in the full six-wave interaction coefficient $C_{123456}$, the LIA contribution, which is of order $\Lambda(k_\mathrm{eff})=\ln(\/k_{\mathrm{eff}}a)$ in the interaction coefficient (\ref{KSW}) directly cancels with the LIA contribution from the six-wave interaction in (\ref{KST}).
The canonical transformation, which is explicitly written in the main text, equation (\ref{eq:C123456}) can be expanded by the means of a small parameter $1/\Lambda(k_\mathrm{eff})$. We expand the denominator, which is just a function of the linear frequency around the $\Lambda(k_\mathrm{eff})$ contribution,

\begin{eqnarray}
\omega_k &\sim  &k^2\left[\Lambda(k_\mathrm{eff})+A\right],\\
\frac{1}{\omega_k} &\sim & \frac{1}{\Lambda(k_\mathrm{eff})}-\frac{A}{\Lambda(k_\mathrm{eff})^2}.
\end{eqnarray}

Using this strategy, we can represent the final six-wave contribution in terms of powers of $1/\Lambda(k_\mathrm{eff})$.  We find in this case that the leading term of order $\Lambda(k_\mathrm{eff})$, which corresponds to the local dynamics drops out, and the next leading order is of order unity.  Moreover, we find that this leading contribution to the six-wave interaction coefficient, which is $\Lambda(k_\mathrm{eff})$ independent is  also $k_\mathrm{eff}$ independent.

Unfortunately, there is no magic cancelations or simplifications in the resulting six-wave coefficient and it
remains
extremely large and complicated  even after
the integrations and the asymptotical limits of large $\Lambda(k_\mathrm{eff})$ taken in this Appendix.
This is why we think that the truncated-LIA model, where the six-wave coefficient is simply the product of
the six wavenumbers, is helpful for understanding the basic properties of Kelvin wave turbulence,
before a more complete model based on the BSE can be attacked. However, we think that the derivations
made in this Appendix will help us to at least establish whether or not the KZ spectra of Kelvin wave
turbulence are local or nonlocal, and a work is underway in this direction.

\section{Appendix B - Integrability of $2$D-LIA}
We show the details of the derivation that the LIA of the BSE yields the $2$D-LIA model (\ref{eq:lia}), i.e. that the
cutoff operation commutes with making the $2$D reduction.
In view of the integrability of the original LIA, this amounts to a proof of integrability of the
$2$D-LIA model (\ref{eq:lia}).

The LIA of the BSE can be written as

\begin{equation}\label{eq:LIAofBSE}
\dot{\bf r}= \beta {\bf r}'\times {\bf r}'',
\end{equation}
where the notation for the differentiation operator $'$ is $d/dl$, and where $l =\left(1+|{\bf w}|^2\right)^{-1/2}$ is the arc length.
 The two-dimensional representation of a vortex line can be described by a vector ${\bf r}=z\hat{z}+{\bf w}$.  The vector ${\bf w} = (x(z),y(z))$ being a function of $z$, orientated in the $xy$-plane.



Applying the chain rule to rewrite all derivatives to be with respect to $z$ (from this point on $'$ will refer to $d/dz$), LIA can be written as
\begin{equation}\label{eq:rdot}
\dot{{\bf r}}=\beta\left[\left(1+|{\bf w}'|
^2\right)^{-3/2}(\hat{z}\times{\bf w}''+{\bf w}'\times{\bf w}'')\right].
\end{equation}
With a little geometrical intuition \cite{S95}, one can show $\dot{{\bf w}}=\dot{\bf r} - (\dot{\bf r}\cdot\hat{z})(\hat{z}+{\bf w}')$. In addition, both ${\bf w}'$ and ${\bf w}''$ are perpendicular to $\hat{z}$ direction, thus, one can represent ${\bf w}'\times {\bf w}''=(({ \bf w}'\times {\bf w}'')\cdot\hat{z})\hat{z}=A\hat{z}$. Then equation (\ref{eq:rdot}) can be reduced to
\begin{equation}\label{eq:wdotfinal}
\dot{\bf w}=\beta \left(1+|{\bf w}'|^2\right)^{-3/2}\left[\hat{z}\times{\bf w}''-A{\bf w}'\right].
\end{equation}

We will now show that equation (\ref{eq:lia}) is equivalent to equation (\ref{eq:wdotfinal}).  First, we must change our representation of $w$ from a complex variable $w(z)=x(z)+iy(z)$ to vector notation ${\bf w}=(x(z),y(z))$.  Equation (\ref{eq:lia}) is equivalent to
\begin{equation}\label{eq:com2vec}
\dot{\bf w}=\frac{1}{2}\hat{z}\times \frac{\partial}{\partial z}\left(
\frac{{\bf w}'}{\sqrt{1+|{\bf w}'|^2}}\right).
\end{equation}
Expanding, keeping track of $\beta$ and applying the vector identity $({\bf a}\times{\bf b})\times {\bf c}=({\bf c}\cdot {\bf a}){\bf b}-( {\bf c}\cdot {\bf b}) {\bf a}$, equation (\ref{eq:com2vec}) can be rewritten as
\begin{equation}
\dot{\bf w}=\beta \left(1+|{\bf w}'|^2\right)^{-3/2}\left[\hat{z}\times {\bf w}''-A{\bf w}'\right].
\end{equation}
This is exactly the same result as Equation (\ref{eq:wdotfinal}). The $2$D-LIA model (\ref{eq:lia}) is equivalent to the LIA of the BSE (\ref{eq:LIAofBSE}) and so the $2$D-LIA model is indeed integrable. We remind, however, that both models (\ref{eq:lia}) and (\ref{eq:LIAofBSE}) can only be used continuously until
the moment when they predict self-crossings of vortex lines (absent in weak turbulence), and therefore the integrability of these models
may only be used for predicting the vortex line motion in between of the reconnection events (see the footnote on page 4).




\end{document}